\RequirePackage{ifpdf}
\documentclass[letterpaper]{JHEP3}
\pdfoutput=1

\usepackage{graphicx}
\usepackage{amsmath}
\usepackage{amsfonts}
\usepackage{epsfig}
\usepackage{slashed}

\usepackage{comment}

\def\bec{\begin{center}}
\def\eec{\end{center}}
\def\beq{\begin{equation}}
\def\eeq{\end{equation}}
\def\bea{\begin{eqnarray}}
\def\eea{\end{eqnarray}}
\def\muhat{\hat{\mu}}
\def\nuhat{\hat{\nu}}

\def\rhohat{\hat{\rho}}

\title{Fermion mass without symmetry breaking}

\author{Simon Catterall\\
Department of Physics, Syracuse University, Syracuse, NY13244, USA \\
E-mail: \email{smcatter@syr.edu}
}

\abstract{We examine a model of reduced staggered fermions in three
dimensions interacting through an $SO(4)$ invariant four
fermion interaction. The model is similar to that considered in a recent paper by Ayyer and
Chandrasekharan \cite{Ayyar:2014eua}.
We present theoretical arguments and 
numerical evidence which support the idea that the system develops a mass gap 
for sufficiently strong four fermi coupling.{\it without}
producing a symmetry breaking fermion bilinear condensate.
Massless and massive phases appear to be separated by a continuous phase transition.}

\begin{document}
\section{Introduction}

In this paper we study a three dimensional lattice theory containing four fermion
interactions invariant under an $SO(4)$ symmetry. The structure of the model is motivated by 
recent work by Ayyar and Chandrasekharan \cite{Ayyar:2014eua} who construct a similar
theory enjoying an $SU(4)$ symmetry. The authors of that earlier work
provide strong evidence that the system is capable of dynamically generating
a fermion mass {\it without} producing a symmetry breaking
bilinear condensate. Such a mechanism of mass generation is novel and potentially may
find application in many areas of physics -- for example it may offer alternative mechanisms
to decouple mirror fermions in efforts to construct lattice chiral gauge theories - see the recent work
in \cite{BenTov:2015gra,You:2014vea} and the recent review \cite{Poppitz:2010at} which
provides a useful set of references to the earlier work in the lattice community.

The $SO(4)$ invariant theory we study
possesses
practical advantages over the $SU(4)$ model. The latter model possesses a sign problem
and must be simulated using a worm algorithm. This limits the size of the system that
can be studied. In contrast the $SO(4)$ model we study in this paper does not suffer
from a sign problem; after one replaces the four
fermion interaction by suitable Yukawa terms, one can show that the Pfaffian that arises after
integrating over the fermions is real, positive definite
and hence can be simulated using a (rational) hybrid Monte Carlo algorithm. 

In the next section we describe the lattice model and its symmetries and the phases expected at strong and weak four fermi
coupling. The use of a {\it reduced} staggered fermion field ensures the absence of a  single site 
mass term and so any $SO(4)$ invariant mass term must be constructed from
products of staggered fields which are neighbor to each
other within an elementary (hyper)cube \cite{vandenDoel:1983mf,Golterman:1984cy}.
In fact the massless staggered quark action is invariant under a set of discrete shift symmetries which
protect the theory against the generation of even these operators. Of course it is entirely possible
that either the $SO(4)$ or shift symmetries could break spontaneously at strong coupling and yield
corresponding condensates and so much of our
paper is devoted to providing numerical evidence to
show that this does not occur.  Instead we argue that the physics at strong coupling
is instead driven
by condensation of the symmetric four fermi operator appearing in the action.  We show
numerical evidence in favor of a phase transition to such a condensed phase for a critical value of
the four fermi coupling. Furthermore, we show that the system develops  a mass in this
condensed phase. The use of reduced staggered fermions for
constructing interesting lattice four fermion models was also emphasized in \cite{Catterall:2013koa, Catterall:2013sto}. The latter work argued for the presence of a Higgs phase in a strongly coupled fermion theory
although it has since been argued that this phase cannot survive the continuum limit \cite{Golterman:2014yha}.

Finally, since both these lattice symmetries 
are also present in four dimensions it is possible that this
phase structure may also be present if the model
is lifted to four dimensions.

\section{Lattice model and symmetries}

Consider a theory of {\it reduced} staggered fermions in three dimensions
whose action contains a single site SO(4) invariant
four fermion term. The action is given by
\begin{equation}
S=\sum_x\sum_\mu \eta_\mu(x) \psi^a(x)\Delta^{ab}_\mu\psi^b(x)-\frac{1}{32}G^2\left(\sum_x \epsilon_{abcd}\psi^a(x)\psi^b(x)\psi^c(x)\psi^d(x)\right)
\end{equation}
where $\Delta^{ab}_\mu\psi^b(x)=\frac{1}{2}\delta_{ab}\left(\psi^b(x+\muhat)-\psi^b(x-\muhat)\right)$ with
$\muhat$ representing unit displacement in the lattice in the $\mu$-direction 
and $\eta_\mu(x)$ is the usual staggered fermion phase $\eta_\mu(x)=\left(-1\right)^{\sum_{i=0}^{\mu-1} x_i}$.
The (single component) reduced staggered
fermions are taken to transform in the fundamental representation of $SO(4)$  according to
\begin{equation}
\psi(x) \to O\psi(x)\end{equation}
with $O$ an arbitrary $SO(4)$ rotation. 
If one employs periodic boundary conditions it is also invariant under the shift symmetry
\begin{equation}
\psi(x)\to \xi_\rho(x)\psi(x+\rhohat)\end{equation}
where the flavor phase $\xi_\mu(x)=\left(-1\right)^{\sum_{i=\mu+1}^{d-1}x_i}$.
In four dimensions these shift symmetries can be thought of as a discrete 
remnant of continuum chiral symmetry \cite{Bock:1992yr}.

These symmetries strongly constrain the possible bilinear terms that can occur in the theory. For example,
a single site mass term of the form $\psi^a(x)\psi^b(x)$
breaks the $SO(4)$ invariance but maintains the shift symmetry\footnote{The usual single site mass term that is possible for a full staggered field
$\overline{\psi}^a(x)\psi^a(x)$ is invariant under both the $SO(4)$ and shift symmetries but this
term is absent for a {\it reduced} staggered field.} while $SO(4)$ invariant
bilinear terms constructed from products of
staggered fields within the unit (hyper)cube generically break the
shift symmetries \cite{vandenDoel:1983mf,Golterman:1984cy}. 
The possible bilinear operators for a reduced staggered fermion that
correspond to Dirac and Majorana masses in the continuum limit are listed below\footnote{We thank Jan Smit for useful conversations concerning the issue of 
reflection positivity with staggered fermions actions}

\begin{eqnarray}
O_1&=&\sum_{x,\mu} m_\mu\epsilon(x)\xi_\mu(x)\psi^a(x)S_\mu\psi^a(x) \\\nonumber
O_2^1&=&\sum_{x,\mu,\nu} m^1_{\mu\nu}\xi_\mu(x)\xi_\nu(x+\muhat)\psi^a(x) S_\mu S_\nu \psi^a(x)\\\nonumber
O_2^2&=&\sum_{x,\mu,\nu} m^2_{\mu\nu} \epsilon(x)\xi_\mu(x)\xi_\nu(x+\muhat)\psi^a(x)S_\mu S_\nu \psi^a(x)\\\nonumber
O_3&=&\sum_{x,\mu,\nu,\lambda} \hat{m}_{\mu\nu\lambda} \xi_\mu(x)\xi_\nu(x+\muhat)\xi_\lambda(x+\muhat+\nuhat)\psi^a(x)S_\mu S_\nu S_\lambda \psi^a(x)
\label{ops}
\end{eqnarray}
where the site parity $\epsilon(x)=\left(-1\right)^{\sum_{i=0}^{d-1}x_i}$ and the
coefficients $m^1_{\mu\nu},m^2_{\mu\nu},\hat{m}_{\mu\nu\lambda}=\hat{m}\epsilon_{\mu\nu\lambda}$ are totally antisymmetric in their indices.  The
symmetric translation operator $S_\mu$ acts  on a lattice field according to
\begin{equation}
S_\mu \psi(x)=\psi(x+\muhat)+\psi(x-\muhat)
\end{equation}
It is straightforward to show that these mass terms are all antisymmetric operators.  As an example consider taking the
transpose of the one link operator $O_1$.
\begin{equation}
O_1^T=\sum_{x,\mu}-m_\mu\epsilon(x)\xi_\mu(x)\left[\psi(x+\muhat)\psi(x)+\psi(x-\muhat)\psi(x)\right]\end{equation}
where we have have used the Grassmann nature of the fermions. Assuming periodic boundary conditions we
can now shift the summation index $x$
\begin{equation}
O_1^T=\sum_{x,\mu}m_\mu\epsilon(x)\xi_\mu(x)\psi(x)\left[\psi(x-\muhat)+\psi(x+\muhat)\right]=O_1\end{equation}
This antisymmetric property is {\it required} of any bilinear operator acting on reduced staggered fermions. Let us
now check the invariance properties of the one link operator under a shift symmetry associated with
the direction $\rho$
\begin{equation}
O_1^{\left(\rho\right)}=\sum_{x,\mu}m_\mu\epsilon(x)\xi_\mu(x)\xi_\rho(x)\xi_\rho(x+\muhat)\psi(x+\rhohat)\left[\psi(x+\muhat+\rhohat)+\psi(x-\muhat+\rhohat)\right]\end{equation}
Again, assuming periodic boundary conditions we can shift the summation index $x$
\begin{equation}
O_1^{\left(\rho\right)}=\sum_{x,\mu}-m_\mu\epsilon(x)\xi_\mu(x)\xi_\mu(\rhohat)\xi_\rho(\muhat)\psi(x)\left[\psi(x+\muhat)+\psi(x-\muhat)\right]\end{equation}
Using the result $\xi_\mu(\rhohat)\xi_\rho(\muhat)=\left(2\delta_{\mu\rho}-1\right)$ we find
\begin{equation}
O_1^{\left(\rho\right)}=O_1-2m_\rho\epsilon(x)\xi_\rho(x)\psi(x)\left[\psi(x+\rhohat)+\psi(x-\rhohat)\right]
\label{onelink}\end{equation}
Thus the expectation value of the
second term on the RHS of eqn~\ref{onelink} must hence vanish {\it unless} the corresponding
shift symmetry is broken spontaneously.
Similar conclusions 
can be obtained for the other multilink mass operators. 
The question of whether these symmetries break spontaneously, which is a prime focus of the
current paper, requires  a careful study of the finite
volume system in the presence of a symmetry breaking external field. Spontaneous symmetry breaking 
is signaled by the presence of a non-zero bilinear condensate as the external source is sent to zero after the
thermodynamic limit is taken. We will show numerical evidence later that this {\it does not} occur in
this model.

Before turning to the auxiliary representation of the four fermi term and the subsequent numerical simulations we
can first attempt to understand the behavior of the theory in the limits of both weak and strong coupling.
At weak coupling one expects that the fermions are massless and there should be
no bilinear condensate since the four fermi term is
an irrelevant operator by power counting. In contrast  the
behavior of the system for large coupling  can be deduced from a strong coupling expansion. 
The leading term corresponds to the static limit $G\to\infty$ in which the kinetic operator is dropped
and the exponential of the four fermi term expanded in powers of $G$.
In this limit the partition function is  saturated by terms of the form
\begin{equation}
Z\sim \left[G^2 \int d\psi^1(x)d\psi^2(x) d\psi^3(x) d\psi^4(x) \psi^1(x)\psi^2(x)\psi^3(x)\psi^4(x)\right]^V
\end{equation}
corresponding to a single site four fermi condensate.
Rescaling the fermion fields by $\sqrt{G}$ then facilitates a calculation of the
the fermion propagator $G_f(x,y)=\left\langle \psi^a(x)\psi^a(y)\right\rangle$  by expanding the
exponential of the kinetic operator each term of which carries a factor of $1/G$.
\begin{equation}
G_f(x,y)\sim
\sum_{{\rm paths}\;P_(x\to y)}
\left(\frac{1}{G}\right)^{2l(P)} \prod_P 
\left(M_{x x_1}\right)^3\left(M_{x_1 x_2}\right)\left(M_{x_2 x_3}\right)^3\left(M_{x_3 x_4}\right)\ldots\left(M_{x_n y}\right)^3
\end{equation}
where $M_{x x+\mu}=\frac{1}{2}\eta_\mu(x)$ and $l(P)$ is the (odd) number of links along the path $P$.
A similar calculation for the bosonic operator $G_b(x,y)=\left\langle \psi^a(x)\psi^b(x)\psi^a(y)\psi^b(y)\right\rangle$ yields an analogous expression
\begin{equation}
G_b(x,y)\sim
\sum_{{\rm paths}\;P_(x\to y)}
\left(\frac{1}{G}\right)^{2l(P)} \prod_P 
\left(M_{x x_1}\right)^2\left(M_{x_1 x_2}\right)^2\left(M_{x_2 x_3}\right)^3\left(M_{x_3 x_4}\right)^2\ldots\left(M_{x_n y}\right)^2
\end{equation}
In both cases the leading term corresponds to paths of minimal length and yields an exponential
behavior for the correlation function with a mass $m\sim 2\ln{G}$.  Notice that
the vanishing of the bosonic correlator
for large separations implies that the corresponding fermion bilinear $<\psi^a(x)\psi^b(x)>$ is also
zero in this limit.

Thus the strong coupling calculation indicates that for sufficiently large $G$
the system should realize a phase in which the fermions acquire a mass without breaking the
$SO(4)$ symmetry. At weak coupling we  have argued that the symmetry is also unbroken but the
fermions are massless.
Clearly there has to be at least one phase transition separating
these two phases and we will indeed provide evidence in favor of this later. But one can
also conceive of a more complicated phase structure with multiple phase transitions
between weak and strong coupling. Indeed 
the lattice theory we have been discussing
is similar to Higgs-Yukawa models employing staggered or naive
fermions that were studied earlier in \cite{Hasenfratz:1988vc,Lee:1989xq,
Stephenson:1988td} although those
models differ crucially from the one considered in this
paper  since they employ a full staggered field and hence allow for
symmetric bilinear mass terms. In particular while
that earlier work provided evidence of symmetric phases at both weak and strong coupling  
a broken phase with a bilinear condensate was also observed at intermediate coupling.
This intermediate phase was separated from its symmetric neighbors by first order
phase transitions. We see no evidence of 
this intermediate phase in our model. 

\section{Auxiliary field representation}

Following the usual Gross-Neveu strategy one can try to rewrite the theory in terms of Yukawa terms
and auxiliary scalars. Specifically,
if we define the antisymmetric fermion bilinear fields
$\Sigma^{ab}=\psi^a\psi^b$ the four fermi term can be
rewritten
\begin{equation}
\frac{1}{16}G^2\sum_x \Sigma^{ab}\tilde{\Sigma}^{ab}\end{equation}
where the dual fermion bilinear is given by $\tilde{\Sigma}^{ab}=\frac{1}{2}\epsilon_{abcd}\Sigma^{cd}$.
Furthermore if we introduce the (anti)self-dual fermion bilinear fields
$\Sigma^{ab}_\pm=\frac{1}{2}\left(\Sigma^{ab}\pm\frac{1}{2}\epsilon_{abcd}\Sigma^{cd}\right)$ and exploit the
Grassmann character of the underlying fermions this can be rewritten as
\begin{equation}
\pm\frac{1}{8}G^2\left(\Sigma_\pm^{ab}\right)^2\end{equation}
Notice that $\Sigma_\pm$ transform in the fundamental representation under
the corresponding $SO_\pm(3)$ subgroup 
of $SO(4)$ (they are singlets under the other $SO_\mp(3)$ symmetry)
\begin{equation}
SO(4)= SO_+(3)\times SO_-(3)\label{so4}\end{equation}  The original action can then be generated by employing an auxiliary
antisymmetric, self-dual boson field 
\beq
\phi_+^{ab}=\frac{1}{2}\left(\phi^{ab}+\frac{1}{2}\epsilon_{abcd}\phi^{cd}\right)
\eeq\noindent
which transforms in the fundamental of $SO_+(3)$ and is a singlet under $SO_-(3)$~\footnote{An similar action based on $\Sigma_-^{ab}$ can be used to construct the theory with $G^2<0$.}
\begin{equation}
\frac{G}{2}\phi_+^{ab}\Sigma_+^{ab}+\frac{1}{2}\left(\phi_+^{ab}\right)^2
\end{equation}
\noindent
The primary
advantage of writing the four fermion term  using auxiliary fields is that
then the fermions appear only quadratically in
the action and so may be integrated out to yield a Pfaffian. In principle
this allows one to simulate the model using the rational hybrid Monte Carlo algorithm \cite{Clark:2004cp}.
However the latter algorithm requires that the Pfaffian be real, positive definite - the theory
must not suffer from a sign problem. In fact we can show
that the self-dual property of the 
auxiliary field guarantees just this property.
Consider the eigenvalue problem associated with the 
fermion operator
\begin{equation}
\left( \eta_\mu\Delta_\mu+\frac{G}{2}\phi_+\right)\psi=\lambda \psi
\end{equation}
Since the operator is real and antisymmetric the eigenvalues $\lambda$ are pure imaginary and
come in pairs $\lambda$ and $-\lambda$. One way to guarantee
a positive Pfaffian is for the eigenvalues to also be doubly degenerate. It is easy to see that this
is the case since the fermion operator is invariant under $SO_-(3)$ transformations. Since $\psi$ transforms
as the $(\frac{1}{2},\frac{1}{2})$ representation under $SO(4)$ it will transform as a spinor under
$SO_-(3)$. Thus each level $\lambda$ will be doubly degenerate. This conclusion has been
checked numerically and guarantees positivity of the Pfaffian.

If we now imagine integrating out the fermions we will generate an effective action for the
auxiliary bosons $\phi^+_{ab}$ of the form
\begin{equation}
S_{\rm eff}=-\frac{1}{2}{\rm Tr}\ln{\left(\eta .\Delta +\frac{G}{2}\phi_+\right)}\end{equation}
This can be expanded perturbatively in $G$:
\begin{equation}
S_{\rm eff}=-\frac{1}{2}{\rm Tr}\ln{\left(\eta .\Delta\right)}+\frac{1}{2}\sum_{n=1}^\infty 
\frac{\left(-1\right)^{n}}{n}\left(\frac{G}{2}\right)^n\;{\rm Tr}\left(M^{-1}\phi_+\right)^n
\end{equation}
where $M_{xy}^{ab}=\delta^{ab}\eta_\mu(x)\Delta^\mu_{xy}$. Rearranging this yields
an expression for the $\rm n^{th}$ order vertex function of the auxiliary field.
\begin{equation}
\Gamma^n(x_1,\ldots x_n)\sim\frac{\left(-1\right)^{n}}{2n}\left(\frac{G}{2}\right)^n\;M^{-1}_{x_1 x_2}M^{-1}_{x_2 x_3}\ldots M^{-1}_{x_n x_1}\end{equation}
The usual Gross-Neveu analysis of symmetry breaking
proceeds from a consideration of the effective potential $V_{\rm eff}(\phi_+)$
which corresponds to the limit in which are the arguments of the
$\phi_+$ are set to a common value $x_i\to x$. But because of the antisymmetry of $M$  
this onsite effective potential vanishes
identically. Actually, one must be a little careful here; the free fermion operator
has an exact zero mode which must
be regulated in order to define $M^{-1}$ 
in a way compatible with the antisymmetry of $M$.  This can be done using
either a multilink mass term of the type discussed earlier or an antiperiodic boundary condition.

Of course a continuum effective potential could arise from any vertex function $\Gamma^n(x_1,\ldots x_n)$
when the arguments $x_i-x_j=O(a)\;i,j=1\ldots n$ so one should be careful in drawing too strong a conclusion
from this observation. Nevertheless it is at least a piece of evidence in support of the idea that
this lattice model may evade the usual Gross-Neveu symmetry breaking scenario. In general the propagator factors are real
but not necessarily positive so that the structure of $\Gamma^n$ is quite complicated. There is
one notable exception to this:
\begin{equation}
\Gamma^2(x,y)=\frac{1}{2}\left(\delta_{xy}+\frac{G^2}{8} \left[M^{-1}_{xy}\right]^2\right)\end{equation}
where we have added in the contribution from the classical action $\frac{1}{2}(\phi_+^{ab})^2$. This
corresponds to a positive definite nonlocal interaction mediated via the massless staggered
fermions. 

\begin{figure}[htb]
\begin{center}
\includegraphics[width=0.8\textwidth]{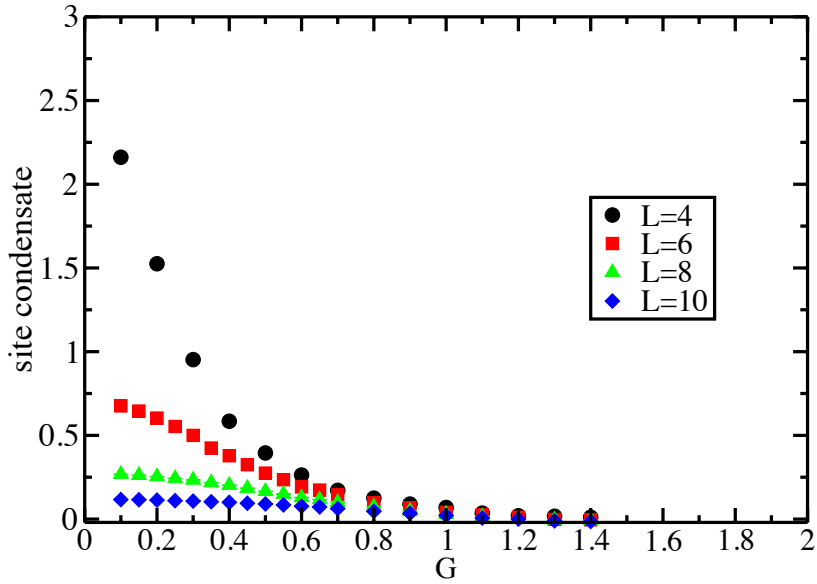}
\caption{\label{bi}$<O_0>$ vs $G$ for $L=4,6,8,10$ and pbc}
\end{center}
\end{figure}\noindent

\begin{figure}[htb]
\begin{center}
\includegraphics[width=0.8\textwidth]{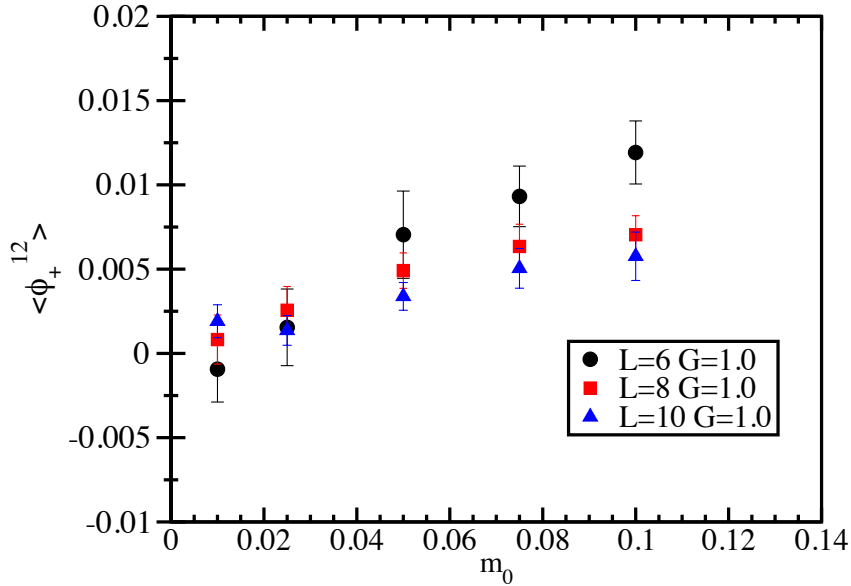}
\caption{\label{mnonzero}$<O_0>$ vs $m_0$ at $G=1.0$ for $L=6, 8,10$ and pbc}
\end{center}
\end{figure}\noindent

\section{Absence of bilinear condensates}

In the previous section we emphasized that the $SO(4)$ and shift symmetries prevent the
finite volume system from generating a non-zero vacuum expectation value for any of the fermion
bilinears operators at finite volume. However these symmetries can in principle break spontaneously in the
infinite volume limit - indeed this is what
is expected to happen in QCD where strong gauge dynamics breaks chiral symmetries
and yields a non-zero chiral condensate. However, in this section we will present evidence that
this does {\it not} occur in this model.

\noindent Consider first a possible site condensate of the form 
\begin{equation}
O_0^{ab}=\left[\psi(x)^a\psi^b(x)\right]_+
\end{equation}
where each index $a,b$ runs over the fundamental representation of $SO(4)$ and the
$+$ subscript signifies as always that we are considering the selfdual component. Since
we are interested in looking for symmetry breaking condensates we
fix the values of these indices. For example figure~\ref{bi}. shows $<O_0^{12}>$ (henceforth
abbreviated to simply $O_0$) from
simulations that
utilize an external source which is
non-zero only for $a=1$ and
$b=2$.
If such an operator were to develop a vacuum expectation value in the thermodynamic limit it would
signal the breaking of $SO(4)\to SO_-(3)\times SO_+(2)$ (again, note that $O_0^{ab}$ is a singlet under
$SO_-(3)$).
To check for such a spontaneous breaking we have added an explicit mass term  of this
form with coupling $m_0=0.1$
to the lattice action and computed the expectation value of the single site condensate
for a range of lattice volumes. Figure~\ref{bi}. shows the condensate
as a function of the four fermi coupling.
Since we employ periodic boundary conditions there is a large contribution to the expectation
value from the near zero mode expected at small $G$. However as can be seen in the plot 
this contribution falls rapidly with increasing
lattice volume. It is also clear that expectation value $<O_0>$
is further suppressed as $G$ increases. This behavior points to
the absence of spontaneous symmetry breaking. This conclusion can be cemented
by examining the behavior of $O_0$ or equivalently $\phi_+^{12}$ at fixed $G$ and for
several lattice volumes as the magnitude of the external source is varied.
Figure~\ref{mnonzero}. shows this quantity  for fixed four fermi coupling $G=1.0$ 
as a function of the external source $m_0$. As expected the condensate
approaches zero on any fixed volume as $m_0\to 0$ but more importantly the
volume dependence is rather weak and indicates a condensate that
is constant or decreasing with increasing volume. This is inconsistent with spontaneous symmetry breaking
which would require that the measured condensate to {\it increase} with volume for
sufficiently small $m_0$. Similar behavior is
seen for all $G$.
Thus the numerical results strongly suggest that the $SO(4)$ symmetry does {\it not} break
spontaneously. 
This is consistent with the results obtained in \cite{Ayyar:2014eua} for
the $SU(4)$ model.
\begin{figure}[htb]
\begin{center}
\includegraphics[width=0.8\textwidth]{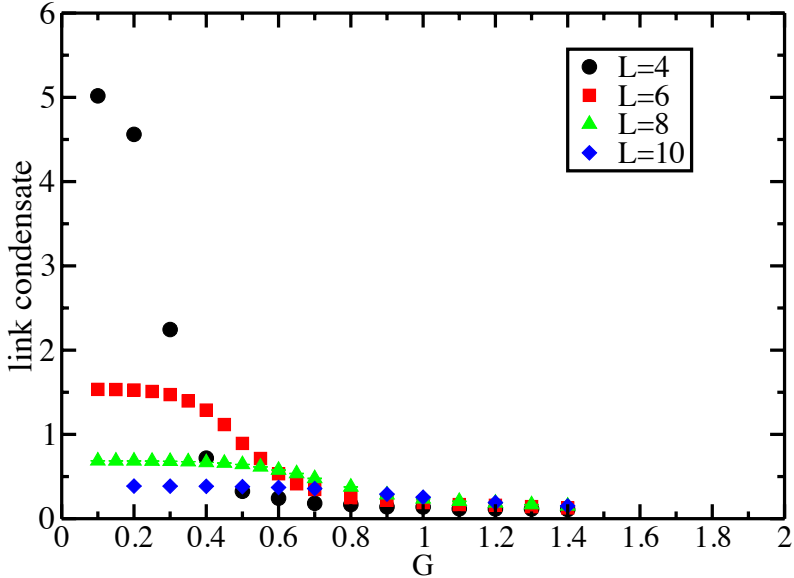}
\caption{\label{link}$<O_1>$ vs $G$ for $L=4,6,8,10$ and pbc}
\end{center}
\end{figure}\noindent
\begin{figure}[htb]
\begin{center}
\includegraphics[width=0.8\textwidth]{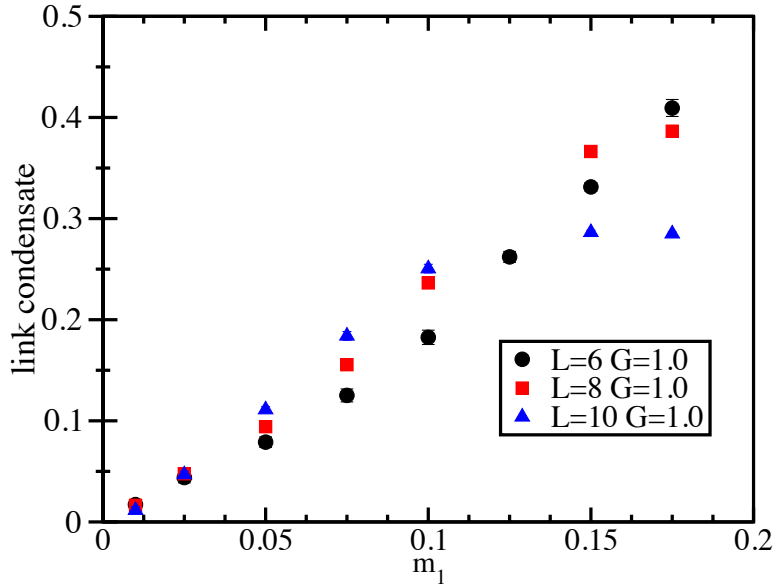}
\caption{\label{m1link}$<O_1>$ vs $m_1$ at $G=1.0$ for $L=6,8,10$ and pbc}
\end{center}
\end{figure}\noindent

Turning to the (hyper)cube mass operators we first analyze the one link operator $O_1$ defined in eqn.~2.4. In our numerical work we consider mass parameters for the form $m_\mu=\left(m_1,m_1,m_1\right)$. 
Again we have examined the possibility that the shift symmetries break spontaneously  by adding
to the action a term of this form coupled to an external field $m_1$. Figure~\ref{link}. shows a plot
of the measured link condensate versus $G$ for a range of lattice volumes
and $m_1=0.1$. As before we employ periodic boundary conditions so that the only source of
shift symmetry breaking lies in $m_1$. In a manner similar to the site operator we see a significant contribution
to the link vev from the (near) fermion zero mode at small $G$ which is rapidly suppressed as the four fermi coupling increases.
We have again checked that the limiting value of $<O_1>$ approaches zero as $m_1\to 0$ in
the large volume limit. Evidence in favor of this is presented in figure~\ref{m1link}. which shows
the link condensate at $G=1.0$ versus $m_1$ for three different lattice volumes. The volume
dependence is rather small and there is no hint of a condensate growing with
volume for small $m_1$ as would be expected if the condensate is to arise from
spontaneous breaking of the shift symmetries. Indeed since the relevant dimensionless parameter governing
a condensate $\Sigma$ is $x=m_1V\Sigma$ the slope of the curve in the case
of spontaneous symmetry breaking should scale like $V$ as $m_1\to 0$
which is clearly not the case with our data.
\begin{figure}[htb]
\begin{center}
\includegraphics[width=0.8\textwidth]{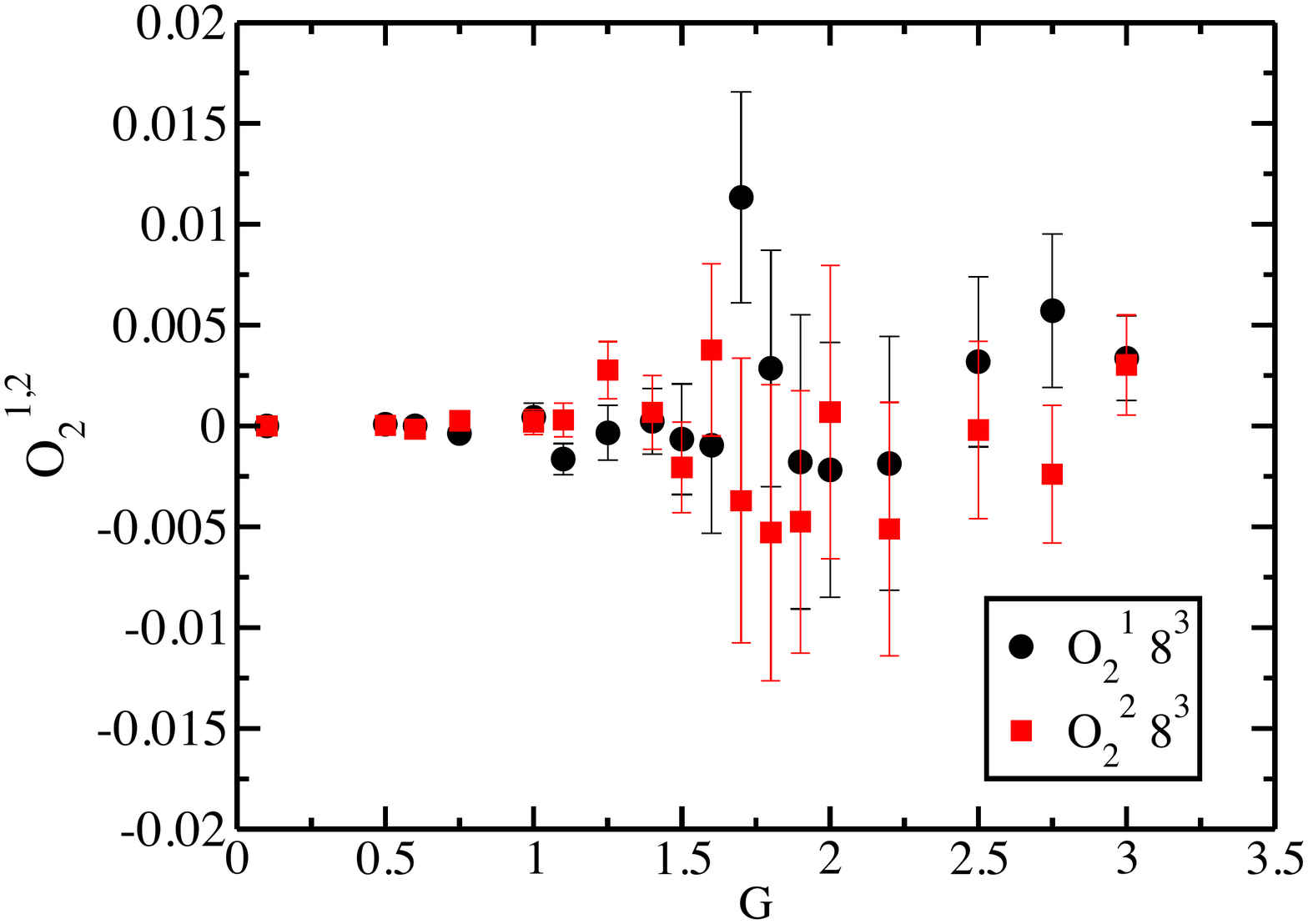}
\caption{\label{sq}$<O_2^1>$ and $<O_2^2>$ vs $G$ for $L=8$ and temporal apbc}
\end{center}
\end{figure}\noindent

\begin{figure}[htb]
\centering
  \includegraphics[width=0.8\textwidth]{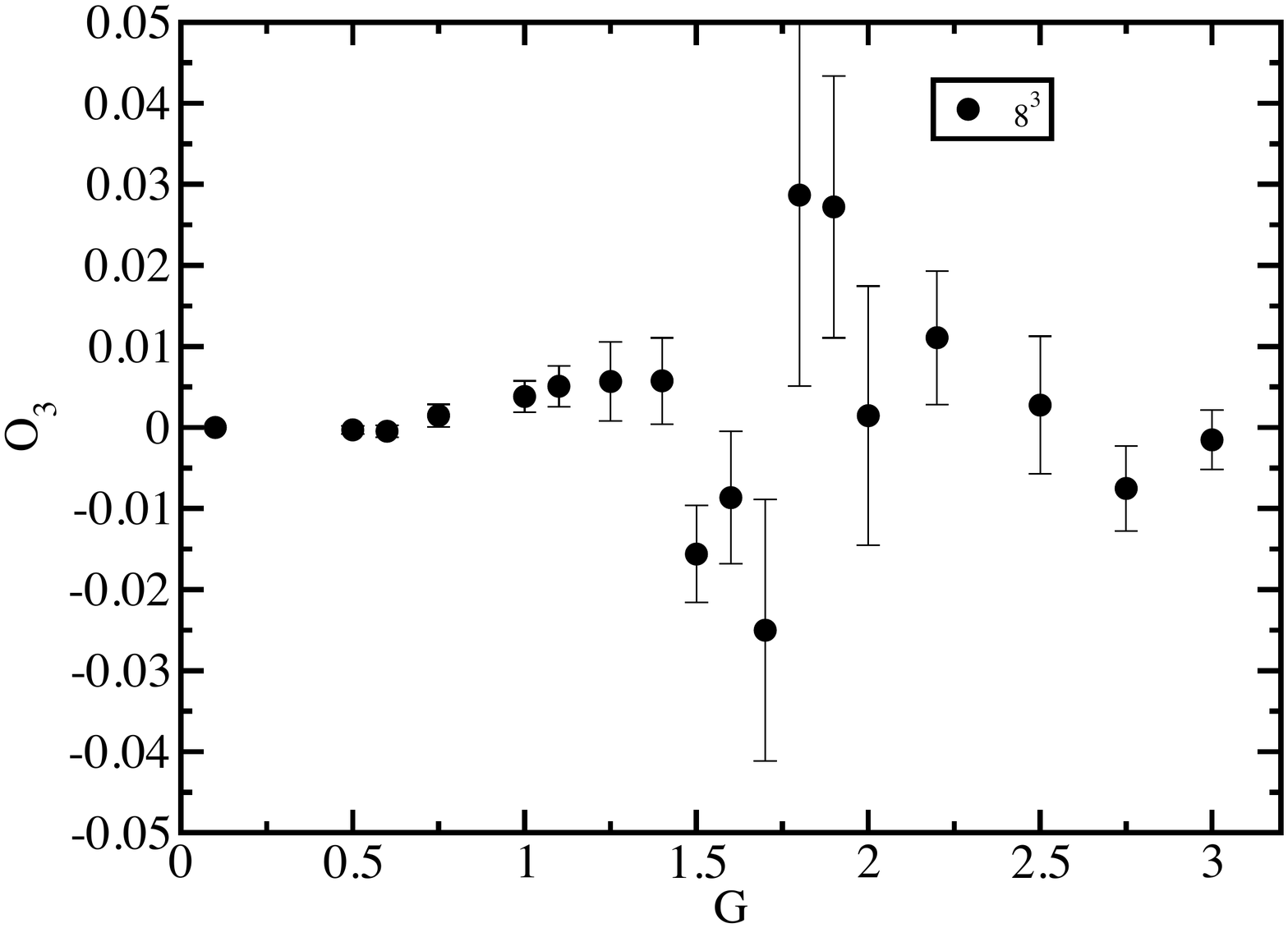}
  \caption{$<O_3>$ vs. $G$ with temporal apbc}
  \label{3link}
\end{figure}

We now turn to the examination of the
two and three link mass operators described earlier. For these observables we show results
obtained with an antiperiodic 
temporal boundary condition and $m_0=m_1=0$.
Such a boundary condition 
explicitly breaks the shift symmetries and yields a fermion mass whose magnitude varies as $O(1/L)$.
It thus provides a suitable way to look for spontaneous breaking of the shift symmetries by monitoring the
vacuum expectation values of all (hyper)cube fermion bilinears simultaneously without introducing 
additional source terms. It has the added merit of
automatically vanishing in the large volume limit. In the next section figure~\ref{lcond}. presents
results for
the $G$ dependence of the one link operator 
which can be seen to be very similar to the previous results shown in figure~\ref{link}.
Using this boundary condition
we have measured the vevs of both types of two link operator and also the three link operator. The two link terms are plotted in figure~\ref{sq}. for an $8^3$ lattice and show that these operators are
statistically consistent with zero over the entire range of coupling $G$. 
The three link $O_3$ operator is shown in figure~\ref{3link}.  also 
for an $8^3$ lattice
and again shows no sign of spontaneous symmetry breaking as the coupling $G$ is varied although
an increase of the statistical errors is clearly visible in the vicinity of $G\sim 2$ which will turn out
to coincide with a rapid increase in the vacuum expectation value of the four fermi operator.
Evidence
for spontaneous breaking of shift symmetry was presented in 
\cite{Cheng:2011ic} for a gauge model in four dimensions. 
One of the order parameters used in that study involves the gauge field and hence has no analog in our
system while the other $L_\mu$ can be written 
\begin{equation}
L_\mu=\sum_{x}\left(-1\right)^{x_\mu} \eta_\mu(x)\psi(x)\left[\psi(x+\muhat)+\psi(x-\muhat)\right]\end{equation}
Using the result $\epsilon(x)\eta_\mu(x)\xi_\mu(x)=\left(-1\right)^{x_\mu}$ and summing
over the index $\mu$ this can be shown to
nothing more than the one link operator examined above and hence contributes nothing new to our
analysis.

\noindent

To summarize it appears that the shift and $SO(4)$ symmetries do {\it not} break spontaneously for
any value of the four fermi coupling. It is not surprising that the symmetries are intact at
weak coupling but it is more surprising that they are not broken as the coupling becomes stronger since
fermionic systems can typically lower their energy by forming condensates. Actually, the existence
of a strong coupled symmetric phase (PMS phase) had been observed before in a variety of lattice
four fermion system - see the review \cite{Poppitz:2010at}. However all these previous studies
found an intermediate symmetry broken phase in which fermion bilinear condensates were
formed. These seem to be absent in this model. Instead the system appears to evolve smoothly
from a massless symmetric phase to the PMS phase with no intermediate symmetry breaking. In the
next section we will provide evidence for this picture.

\begin{figure}[htb]
\begin{center}
\includegraphics[width=0.8\textwidth]{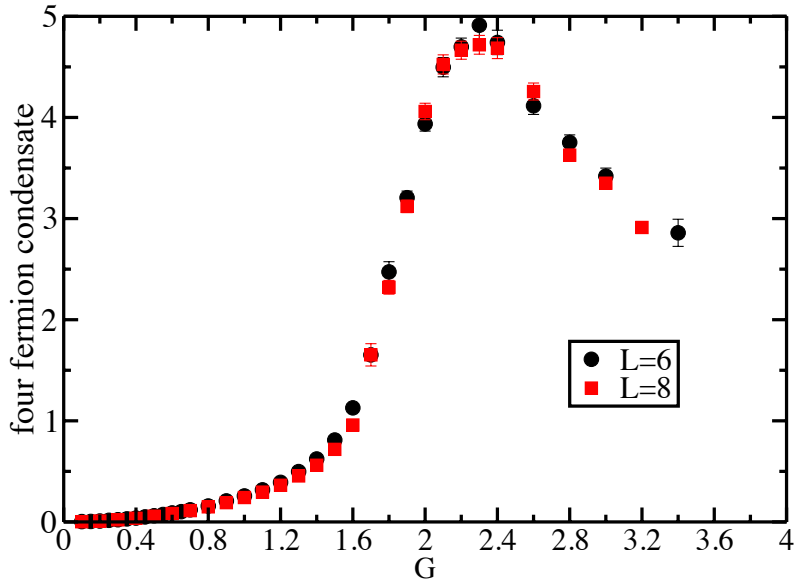}
\caption{\label{four}$\frac{1}{4!}<\epsilon^{abcd}\psi^a\psi^b\psi^c\psi^d>$ vs $G$ with temporal apbc}
\end{center}
\end{figure}\noindent

\section{Phase structure}

As discussed in the previous section we now set the external sources $m_0$
and $m_1$ to
zero and instead use a thermal boundary condition to regulate the would be fermion zero mode in the
weak coupling phase.
To probe the phase structure of the theory we plot the expectation value of the four fermion
operator $O_4=\frac{1}{4!}\epsilon^{abcd}\psi^a\psi^b\psi^c\psi^d$
as a function of the coupling $G$.\footnote{Each ensemble used for the phase structure
analysis at fixed $G$ and $L$
consists of 1000 configurations each separated by 10 HMC trajectories.} 

\begin{figure}[htb]
\begin{center}
\includegraphics[width=0.8\textwidth]{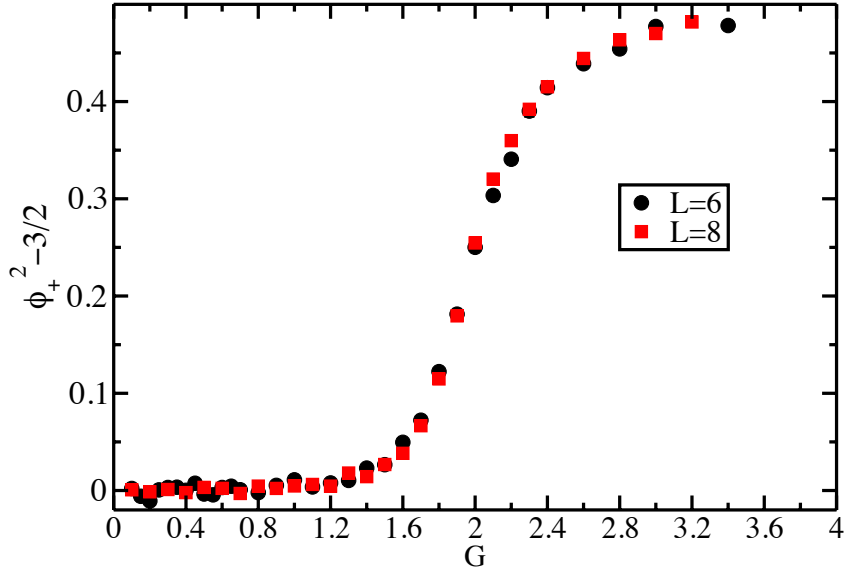}
\caption{\label{plus}$<\phi_+^2>$ vs $G$ with temporal apbc}
\end{center}
\end{figure}

\begin{figure}[htb]
\centering
\begin{minipage}[b]{.45\textwidth}
  \includegraphics[width=1.0\linewidth]{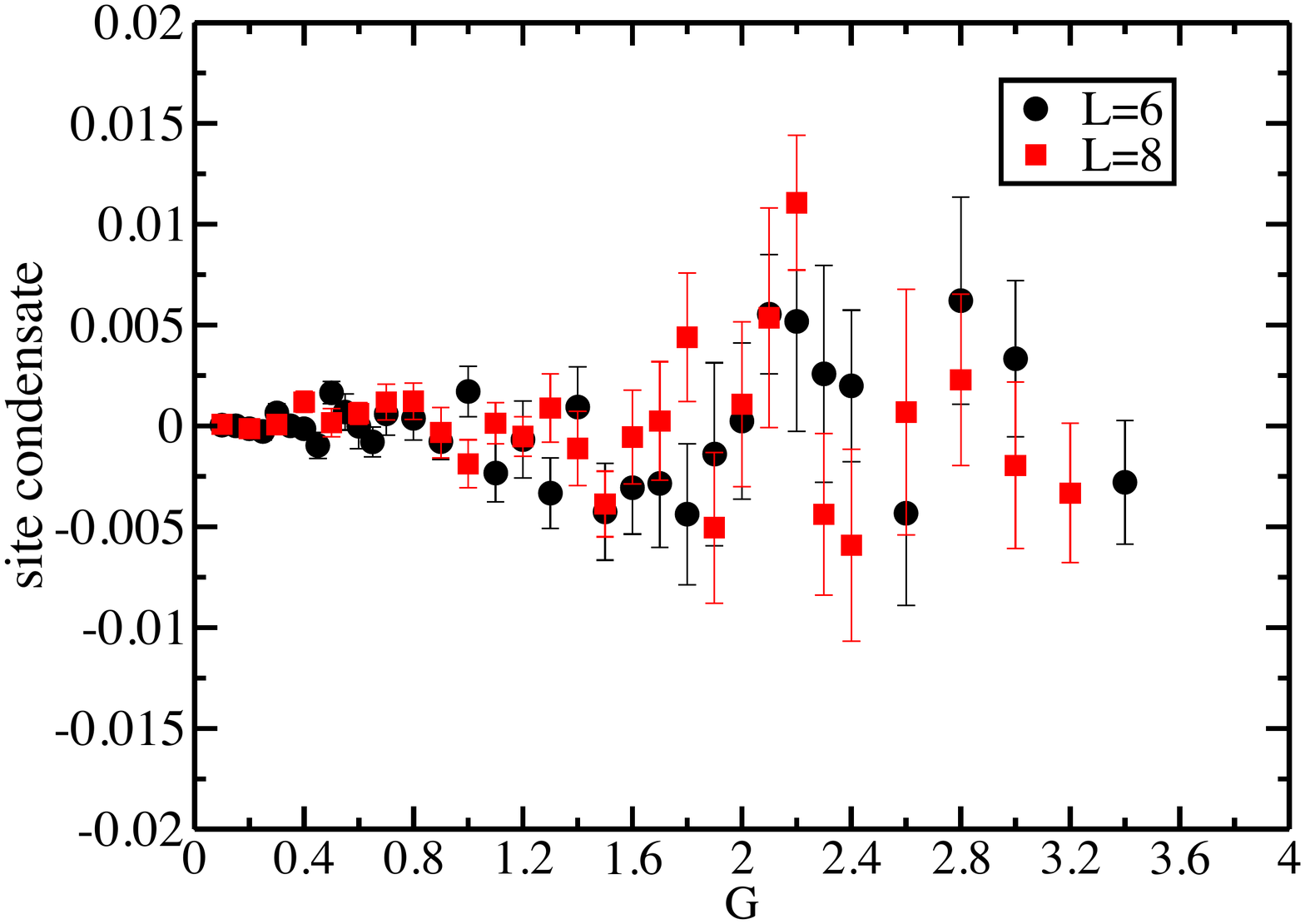}
  \caption{$O_0$ vs. $G$ with temporal apbc}
  \label{bilinear}
\end{minipage}
\quad
\begin{minipage}[b]{.45\textwidth}
  \centering
  \includegraphics[width=1.0\linewidth]{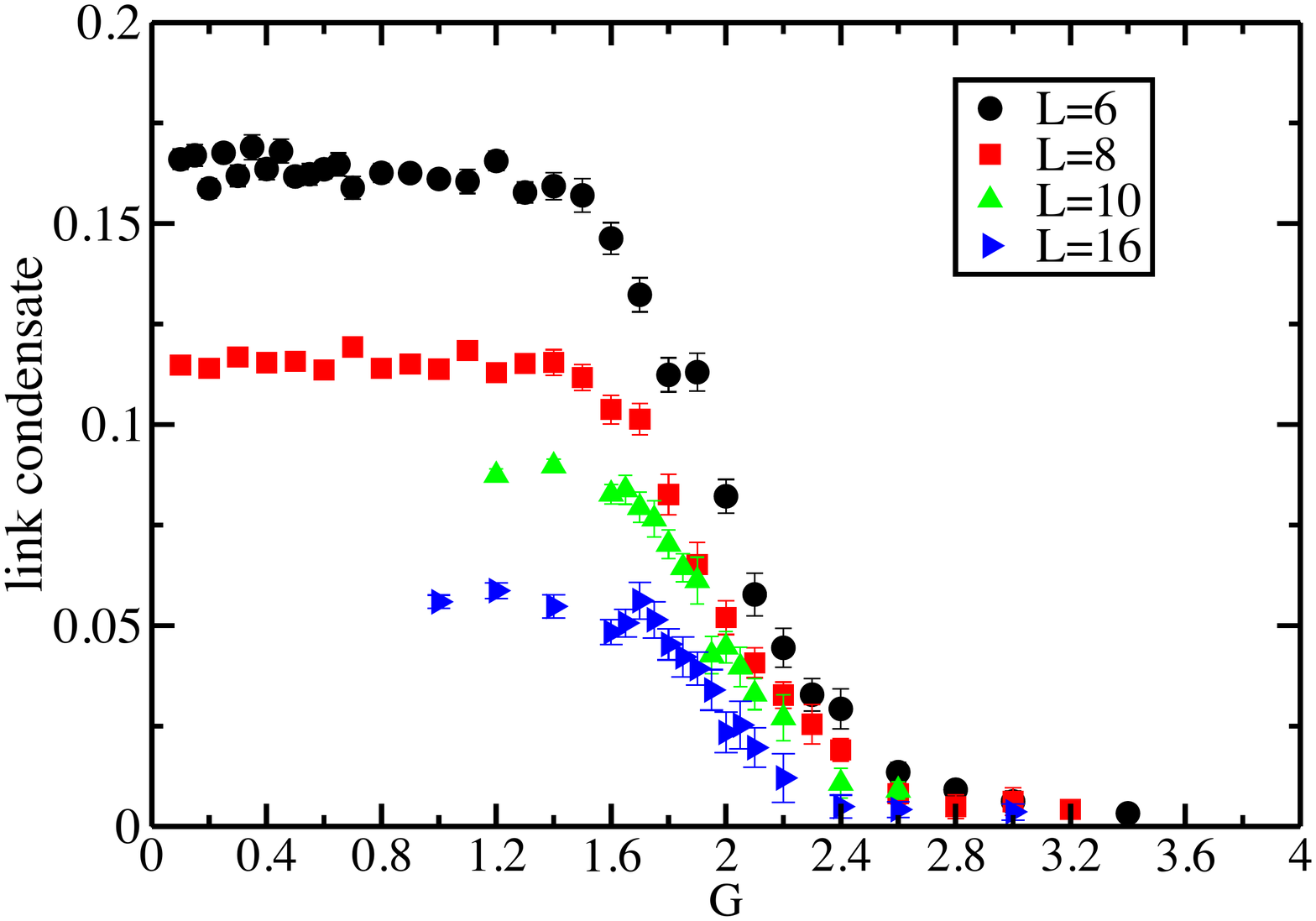}
  \caption{$O_1$ vs. $G$ with temporal apbc}
  \label{lcond}
\end{minipage}
\end{figure}

The expectation value of this operator is plotted in figure~\ref{four}. for two volumes $6^3$ and $8^3$ and shows a peak around $G\sim 2$.
A  related behavior is seen in the square of the auxiliary field $<\frac{1}{2}\phi_+^2>$ 
shown in figure~\ref{plus}. (actually we plot the quantity $<\phi^2_+>-\frac{3}{2}$ which vanishes at $G=0$).
\noindent
The behavior of $\phi_+^2$ is consistent with the usual picture that the minimum of
the effective potential for the auxiliary field
moves away from the origin at strong coupling. What is
different here is that while we observe a rapid increase of $<\phi_+^2>$ for
some $G$ we find good evidence that $<\phi_+>=0$ for all $G$.  This is made clear
in figure~\ref{bilinear}. which plots the expectation value of the single site fermion bilinear (the equivalent plot
for $\phi_+$ itself is very similar).
\noindent
This behavior is quite different from that seen in the standard Gross-Neveu scenario and indicates that
the mechanism driving any phase transition in the model is quite different.

Notice that the thermal boundary conditions do force a non-zero value for the link operator in figure~\ref{lcond}. at
weak coupling but this is suppressed as the four fermi condensate is formed. Furthermore, notice that
the value of this link condensate is approximately constant until the phase transition associated with the
formation of the four fermion condensate is reached - there is no sign of an intermediate broken phase which would
be associated with an enhanced bilinear condensate. The non-zero value of the link operator at
weak coupling is best thought of as arising from an effective link mass term originating in the thermal
boundary conditions. Indeed, the plot makes it clear that the value of the link operator at weak coupling goes to zero
like $1/L$ as the lattice size increases. Thus in the thermodynamic limit our numerical work points to
a vanishing link condensate for
all values of the coupling $G$.

\begin{figure}[htb]
\begin{center}
\includegraphics[width=0.8\textwidth]{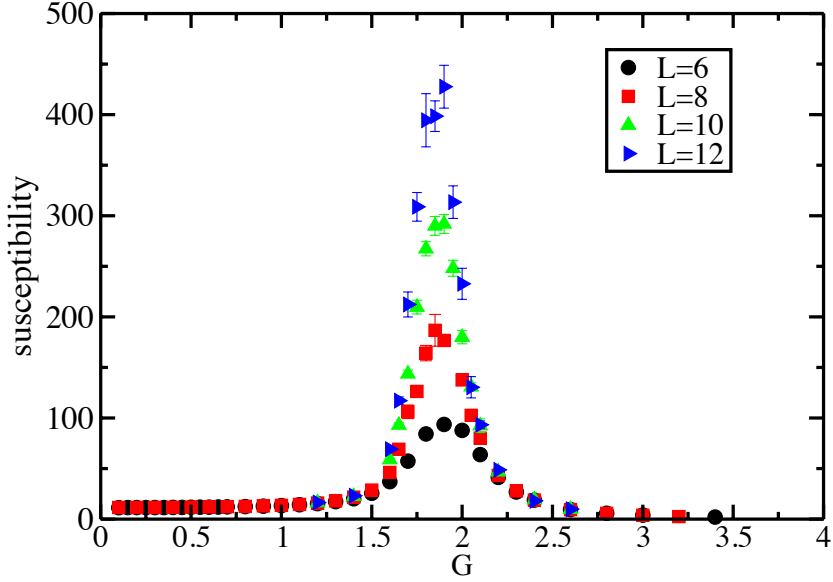}
\caption{\label{sus}$\chi$ vs $G$ for $L=6,8,10,12$ for temporal apbc}
\end{center}
\end{figure}

\begin{figure}[htb]
\begin{center}
\includegraphics[width=0.8\textwidth]{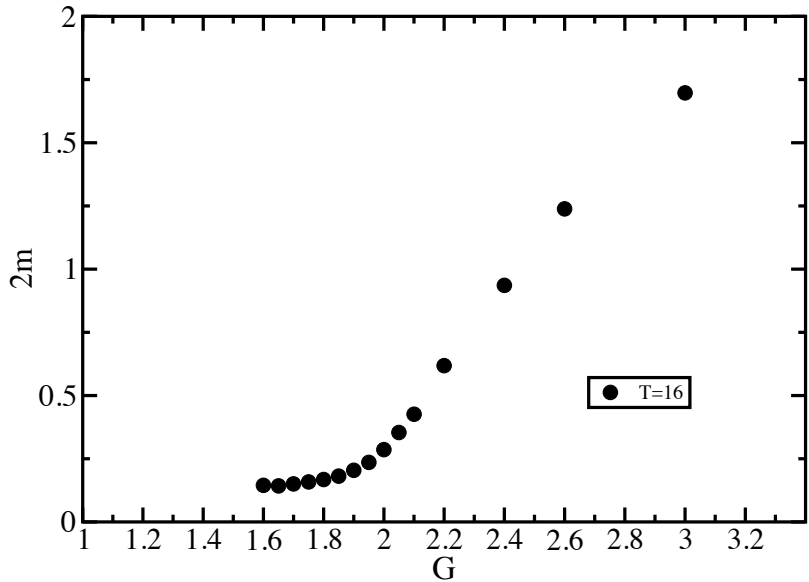}
\caption{\label{mass}Fermion mass vs $G$ for temporal apbc}
\end{center}
\end{figure}

Next we turn to an observable which shows clearly the presence of a single phase
transition at strong coupling. This is 
the susceptibility $\chi$ given by
\begin{equation}
\chi=\frac{1}{L^3}\sum_{a,b,x\ne y} \left\langle \psi^a(x)\psi^b(x)\psi^a(y)\psi^b(y)\right\rangle
\end{equation}
This is plotted in figure~\ref{sus}. for lattice volume $L=6^3,8^3,10^3,12^3$
and shows a single growing peak around $G_c\sim 1.9$ separating the weak and strong coupling regimes. 
The current lattices
are too small for the reliable extraction of critical exponents using finite size
scaling and we leave that for future work. Nevertheless figure~\ref{sus}.
makes it clear that the peak height is growing much more slowly than the volume of the
lattice as would
be expected for a system undergoing a first order phase transition. Hence our current
results favor a continuous phase transition.  
\noindent
\begin{figure}[htb]
\begin{center}
\includegraphics[width=0.85\textwidth]{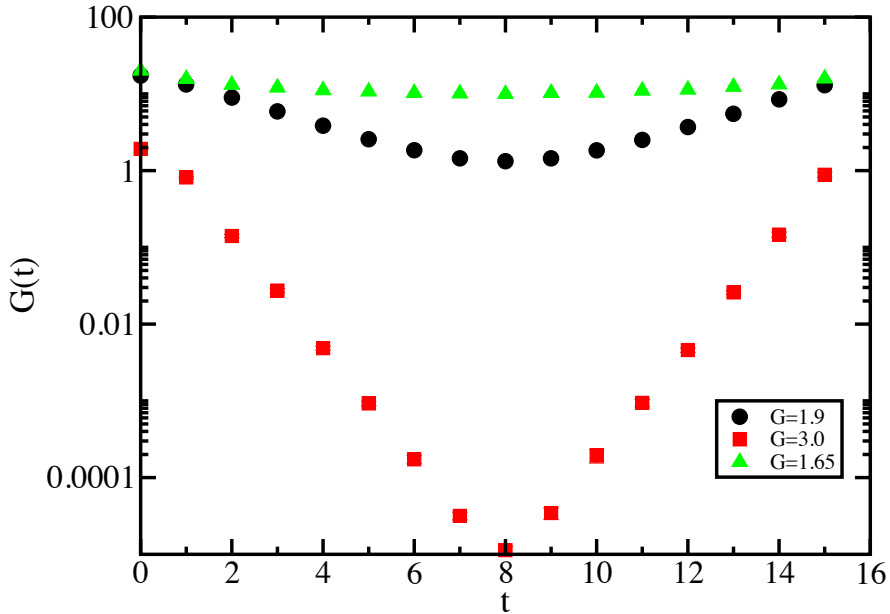}
\caption{\label{corr} $G(t)$ vs $t$ for $G=1.65$, $G=1.9$ and $G=3.0$}
\end{center}
\end{figure}\noindent
Finally we turn to the question of the mass of the fermions. Figure~\ref{corr}. show a plot of the following
spatially averaged four fermion correlation function computed on a $8^2\times 16$ lattice
\begin{eqnarray}
G(t)&=&\sum_{x,y,a,b}\left\langle \psi^a(x,t)\psi^b(x,t)\psi^a(y,0)\psi^b(y,0)\right\rangle\\\nonumber
&=&\sum_{x,y,a,b}
\left\langle\psi^a\left(x,t\right)\psi^b\left(y,0\right)\right\rangle\left\langle\psi^b\left(x,t\right)\psi^a\left(y,0\right)\right\rangle-
\Bigl\langle\psi^a\left(x,t\right)\psi^a\left(y,0\right)\Bigr\rangle^2
\end{eqnarray} 
This quantity when integrated over Euclidean time yields the
susceptibility considered earlier.
We show this correlator in three regimes: weak coupling $G=1.65$, strong coupling $G=3.0$ and close to the phase transition $G=1.9$. As expected the measured mass
is small at weak coupling
but becomes large $O(1)$  at strong four fermi coupling. 
Except for perhaps the $G=1.9$ data the linearity of the log plots is consistent with the correlators
being dominated by a single state for large time separations. Using this information we have
fitted the correlation function to a simple cosh form to extract an estimate of the mass of the 
$\psi^a\psi^b$ state - see the plot in fig~\ref{mass}. Notice that 
the small mass visible in the weak coupling phase is
consistent with the magnitude of the measured one link condensate induced via the use of the
temporal antiperiodic boundary condition. We hence expect the mass to vanish in the thermodynamic
limit. 

\noindent

To summarize our simulations indicate
that the model exhibits a two phase structure; at weak coupling the fermions are
massless while at strong coupling they
acquire a dynamical mass and  the system develops a four fermion condensate. A single continuous
phase transition separates these two regimes but unlike the usual Gross-Neveu scenario there is no
local order parameter - all fermion bilinears vanish at all values of the four fermi coupling
and the  mass generation is {\it not} associated with a spontaneous breaking of
lattice symmetries.

\section{Conclusions}

In this paper we have studied
a three dimensional lattice theory
of reduced staggered fermions coupled through a $SO(4)$ invariant four fermion interaction.
We have put forward theoretical arguments based on exact lattice symmetries together
with the results of numerical simulations to make the case that
bilinear fermion condensates do not form in the theory for any value of the four fermi coupling $G$. This
contrasts sharply from previous lattice studies of four fermion theories - see the reviews \cite{Poppitz:2010at, Hands:1997xv}
and references therein. At weak coupling we observe a conventional paramagnetic (PMW) phase
while for large values of the
coupling a strongly coupled (PMS) phase is observed characterized by a symmetric
four fermion condensate. Both of these phases had been seen before - see
\cite{Golterman:1992yha, Hasenfratz:1988vc}. The existence of the PMS phase and the resulting
gapped spectrum is likely a 
robust feature  of many strong coupling systems. The new aspect of both our
work and that of \cite{Ayyar:2014eua} is that unlike the earlier studies we find no evidence of
a conventional symmetry broken phase at intermediate coupling -- the weak and strong coupling regimes
appear to be separated by a single phase
transition. Furthermore this transition appears to be continuous in contrast to the first order transitions
separating the PMS and PMW phases from the intermediate broken phase. This latter feature
is potentially very important; it suggests that it might be possible to construct a gapped {\it continuum} theory 
close to the phase transition. This should be contrasted with the older work where the masses
generated in the PMS are necessarily cut-off scale. If it is indeed possible
to build a continuum theory with a finite mass gap
close to this putative continuous phase transition it would
constitute a  new mechanism for mass generation which is not tied to the formation
of a symmetry breaking condensate.

The essential ingredients that appear to be important in constructing strongly coupled
lattice fermion models that generate a mass gap
without symmetry breaking are rather restrictive and include 
\begin{itemize}
\item The model must not admit symmetric fermion bilinears. If these are allowed the
system will most likely choose to condense such operators. 
\item The exact symmetries, the dimension of the fermion representation and the structure of
the multifermion operator appearing in the action are intimately related;
in the case of a four fermion term this seems to require
(reduced staggered) fermions appearing in the fundamental representation of either an $SU(4)$ or
$SO(4)$ symmetry group.
\item The dynamics should be such as to render a symmetric four
fermion condensate energetically favorable
in comparison to a symmetry breaking bilinear condensate. While it is plausible that the ground state of a strongly interacting system
breaks the minimal number of exact symmetries this is ultimately a dynamical and non-perturbative question. 
\end{itemize}

The problem of using quartic interactions to gap out fermions without breaking symmetries has also 
received a lot of attention in recent years
from the condensed matter community in the context of topological insulators and superconductors.
It is intriguing that the counting of fermions required in those Hamiltonian constructions matches
closely with the counting of (reduced) staggered quarks in three and four dimensions 
\cite{Fidkowski:2009dba,Morimoto:2015lua,You:2014vea,BenTov:2015gra}. The staggered quark symmetries
seem to play a similar role to time reversal symmetry in the CMT constructions.  Indeed, It is 
possible that the model discussed here is related to that discussed in \cite{Slagle:2014vma}. 

Clearly there is much work to be done in elucidating the nature of the phase diagram and the precise
connection, if its exists,
to these condensed matter systems. We regard the current simulations as a ``proof of principle" and
to establish definitive results and reliable estimates for critical exponents
one will need to simulate much larger lattices. We hope to report on such
results in the near future.  The question of the continuum limit is also of primary importance. Since
$G$ is an irrelevant operator it should flow to zero in the IR and the
physics of the massless phase will be dominated by the trivial fixed point at $G=0$. Any fixed
point at $G=\infty$ would correspond to zero correlation length and have no continuum counterpart.
The interesting question is whether a new continuum limit can be
obtained by tuning $G$ towards $G_c$ assuming the transition is indeed continuous. The nature of this continuum theory
is course very interesting; given the antisynmmetry of the
fermion operator it is natural to consider constructing a continuum theory in which the reduced staggered
field gives rise to four Majorana fermions.
However, a simple transcription of the lattice action will not work - the required four fermion term vanishes identically 
\begin{equation}
\epsilon_{abcd}\psi^{a}_\alpha C_{\alpha\beta} \psi^b_\beta\psi^{c}_\gamma C_{\gamma\delta} \psi^d_\delta=0\end{equation}

Perhaps the most pressing question though, is whether
the phase structure we have seen in three dimensions 
survives to four dimensions. The symmetries prohibiting bilinear terms
remain the same  so it is an open question. 
Whether
any phase transition between weak and strong coupling
remains continuous is another question; the usual universality
arguments would place
this model in the same class as Higgs-Yukawa systems \cite{Hasenfratz:1991it} and so one would
expect that
any phase transition in four dimensions should either be first order or
have mean field  critical exponents. Any evidence to the contrary would be very
exciting!

\acknowledgments SMC is supported in part by DOE grant
DE-SC0009998 and would like thank Poul Damgaard, Anna Hasenfratz, David Schaich, Uwe Wiese,
Rohana Wijewardhana,
Cenke Xu and especially Shailesh Chandrasekharan for 
useful discussions.  SMC would also like to acknowledge
the Aspen Center for
Physics (NSF grant PHY-1066293) and the Kavli Institute for Theoretical Physics
(NSF PHY11-25915) for support during the completion of this
work.
Numerical simulations were performed at Fermilab using USQCD resources.

\bibliographystyle{ieeetr}
\bibliography{so4refs}

\end{document}